\documentclass[reprint, superscriptaddress, amsmath,amssymb, aps, prx, twocolumns, citeautoscript]{revtex4-2}
\usepackage{graphicx}
\usepackage{dcolumn}
\usepackage{bm}
\usepackage[utf8]{inputenc}
\usepackage{placeins} 
\usepackage{appendix}

\begin{document}
\setcitestyle{super} 
\title{An rf Quantum Capacitance Parametric Amplifier}

\author{A.~El~Kass}
\affiliation{ARC Centre of Excellence for Engineered Quantum Systems, School of Physics, The University of Sydney, Sydney, NSW 2006, Australia}
\affiliation{School of Electrical and Information Engineering, The University of Sydney,
Sydney, NSW 2006, Australia}
\author{C.~T.~Jin}
\affiliation{School of Electrical and Information Engineering, The University of Sydney,
Sydney, NSW 2006, Australia}

\author{J.~D.~Watson}
\affiliation{Microsoft Quantum Lab West Lafayette, West Lafayette, Indiana, USA}

\author{G.~C.~Gardner}
\affiliation{Microsoft Quantum Lab West Lafayette, West Lafayette, Indiana, USA}

\author{S.~Fallahi}
\affiliation{Microsoft Quantum Lab West Lafayette, West Lafayette, Indiana, USA}

\author{M.~J.~Manfra}
\affiliation{Microsoft Quantum Lab West Lafayette, West Lafayette, Indiana, USA}
\affiliation{Department of Physics and Astronomy, Purdue University, West Lafayette, Indiana, USA}

\author{D.~J.~Reilly$^\dagger$}
\affiliation{Microsoft Quantum Lab Sydney, The University of Sydney, Sydney, NSW 2006, Australia}
\affiliation{ARC Centre of Excellence for Engineered Quantum Systems, School of Physics, The University of Sydney, Sydney, NSW 2006, Australia}
\date{\today}
\begin{abstract}
We demonstrate a radio-frequency parametric amplifier that exploits the gate-tunable quantum capacitance of an ultra-high mobility two dimensional electron gas (2DEG) in a GaAs heterostructure at cryogenic temperatures. The prototype narrow-band amplifier exhibits a gain $> 20$~dB up to an input power of $-66$~dBm (1-dB compression), and a noise temperature $T_N\sim$~1.3~K at $\sim$~370~MHz. In contrast to superconducting amplifiers, the quantum capacitance parametric amplifier (QCPA) is operable at tesla-scale magnetic fields and temperatures ranging from milli-kelvin to a few kelvin. These attributes, together with its low power (microwatt) operation when compared to conventional transistor amplifiers, suggest the QCPA may find utility in enabling on-chip integrated readout circuits for semiconductor qubits or in the context of space transceivers and radio astronomy instruments.


\end{abstract}
\maketitle


\section{Introduction}

Technologies spanning quantum computing\cite{Review}, radio astronomy\cite{radioast,radioast2}, space communications\cite{space}, particle physics detectors\cite{particle}, and medical imaging systems\cite{MRI} are united in a common quest to extract maximum information from inherently weak signals, ensuring that every available photon contributes to a measurement. Central to this quest are the attributes of the first stage amplifier, which, in terms of its gain, noise, bandwidth, and power dissipation, largely dominates the performance of the readout sub-system or receiver. In quantum computing, for instance, readout plays a key role in qubit measurement, enabling error correcting protocols, qubit bring-up and tuning, as well as yielding the result of a quantum algorithm. Any improvement in the performance of the readout sub-system, in terms of its fidelity, speed, and efficient use of hardware resources directly impacts the entire quantum computing platform \cite{Wallraff_PRApp,Review,Hornibrook_mux,Schaal_NatEl}.



    \begin{table*} [ht] 
\begin{center}
\begin{tabular}{ c c c c | c | } 
 \cline{5-5}
 \textbf{Specs} & \textbf{JPA / TWPA} & \textbf{HEMT} & \textbf{Junction Varactor Paramp} & \textbf{QCPA} \\ 
 \hline
 Gain & 20--30~dB & 40~dB & 20--30~dB & 20--30~dB \\ 
 Lowest Frequency & $\approx$1~GHz & 100s MHz & 100s~MHz & 100s~MHz \\ 
 Bandwidth & Octave & Decade & Octave & Narrow (1--2~MHz) \\ 
 \hline
 Noise Temperature & 100s~mK & 1--2~K & 10~K & 1.29~K \\ 
 Physical Temperature & $<$1~K & $<$20~K & 300--4~K & $<$10~K \\ 
 Power Consumption & Order of nW / $\mu$W & $>$20 mW & $>$20 mW & Order of $\mu$W \\  
 Gain Compression & -100~dBm / -60 dBm & -10~dBm & -10~dBm & -66~dBm \\  
 \hline
 Magnetic Field Compatible & No & Yes & Yes & Yes \\ 
 Linearity & Moderate & High & Moderate & Moderate \\ 
 Directional & No, uses circulator & Yes & No, uses circulator & No, uses circulator \\ 
 Technology & Superconductor & III-V Heterostructure & Abrupt PN junction & III-V Heterostructure \\ \cline{5-5}
\end{tabular}
\end{center}
\caption{Low-noise amplifier specifications representing common state-of-the-art technologies including the historical varactor paramp and the QCPA reported here.}
\label{tab:AmpComp}
\end{table*}

Since the first stage amplifier in the readout chain has dramatic impact on the overall system performance, significant efforts have been made to develop amplifiers that add only the bare minimum noise required by quantum mechanics \cite{Caves,DevScholNature}. To this end, the fluctuation-dissipation theorem \cite{Kubo} intuitively implies that one should look to non-dissipative devices exhibiting superconductivity as a means of realizing amplifiers with the lowest noise. Indeed, various families of superconducting amplifiers have been developed over the last decades including those based on superconducting quantum interference devices (SQuIDs) \cite{Clarke,Slug}, bi-stable junctions \cite{irfan}, and parametric approaches \cite{castellanos2007widely,macklin2015near,yamamoto2008flux,white2015traveling,malnou2020three,eom2012wideband,vissers2016low,JoFET}. Despite the success of these superconducting implementations, challenges remain, including their limited frequency of operation, low input-power compression, and difficulty operating in even modest magnetic fields. In contrast, semiconductor amplifiers, such as those based on high electron mobility transistors (HEMTs) \cite{HEMTs} or SiGe heterojunction bipolar transistors (HBTs)\cite{SiGe}, address these challenges but at the cost of adding much more noise and significant power dissipation.

Here, we report a new kind of rf parametric amplifier based on the modulation of a low-loss `quantum capacitance' associated with an ultra-high mobility two-dimensional electron gas (2DEG) in a GaAs/AlGaAs semiconductor heterostructure grown using molecular beam epitaxy (MBE). With an rf pump voltage applied to a surface gate, the density of the 2DEG is modulated, leading to an oscillating quantum capacitance and parametric gain as the 2DEG is periodically depleted and accumulated. Very recently, the quantum capacitance of a quantum dot was also used to yield a small parametric gain based on the tunneling of electrons \cite{Fernando}. In contrast, 2D heterostructures in the ballastic transport regime can exhibit a much larger modulated quantum capacitance (relative to the total capacitance) leading to large gain operation. In such systems dissipation can be suppressed dramatically\cite{Manfra}, with mean free paths in excess of 100s of microns now relatively commonplace. In the limit where dissipation is reduced to zero the quantum capacitance parametric amplifier (QCPA) acts as a dual of the well-known Josephson parametric amplifier \cite{Fernando}. 

Short of this limit, the presence of moderate dissipation in the QCPA still yields specifications that fill a gap between superconducting para-amps and semiconductor approaches based on cryo-cooled transistors. Operating in the UHF band typically used for semiconductor qubit readout and radio astronomy (0.3 - 3 GHz), the QCPA exhibits gain of $>$ 20 dB, 1 dB compression at input power of -60 dBm, and in the prototype presented here, a noise temperature comparable to state-of-the-art transistor amplifiers based on HEMTs ($\sim$ 1 K). Unlike HEMT or HBT amplifiers however, where the lowest achievable noise is constrained by their (milli-watt) self-heating \cite{SelfHeating}, our prototype QCPA operates at milli-kelvin temperatures with micro-watt power dissipation, suggesting that its noise can likely be further reduced by suppressing ohmic-losses in the gates, 2DEG, and contacts. Via the use of superconducting contacts\cite{supersemi}, for instance, or by switching from GaAs to InAs materials that do not exhibit a Schottky barrier \cite{Mead,InAs_SC} it is possible that dissipation and thus self-heating can be lowered further still. Given our amplifier can be realized in the same 2DEG platform that forms the basis of various semiconductor qubit proposals \cite{SpinsRMP,Spinsgermanium,ManfraInAs}, we anticipate the QCPA may find utility in monolithic integration of readout detectors and amplifiers that add gain to the signal before coming off the chip. 



\section{Background}
The use of variable capacitors to construct parametric amplifiers is, of course well-known, with a rich history dating back to the late 1950's \cite{herrmann1958noise}, where their use in radio astronomy \cite{de1962parametric,jelley1963potentialities} and radar systems was fundamental \cite{adler1961electron}. The early `paramps' relied on diode varactors to provide the variable reactance responsible for the parametric amplification \cite{mumford1960some}. Since the diode is reverse-biased however, thermal noise (rather than shot noise) dominates their performance, with most low-noise applications requiring cryogenic cooling in liquid helium. At such temperatures the performance of varactor-based paramps was theoretically projected to rival the maser \cite{heffner1958minimum,heffner1959solid}. In practice however, pumping at the required power levels of 10s of mW at microwave frequencies causes the junction temperature to rise  locally, well above the ambient temperature of the liquid helium coolant \cite{garbrecht1966noise}. Consequently, this self-heating yielded noise temperatures in the vicinity of 10~K \cite{blake1963helium}. Eventually, heterostructure-based FET devices arrived (in the 1980s) and mostly superseded the original varactor paramps by offering an affordable low-noise alternative \cite{weinreb1980low} with the simplicity of dc bias rather than microwave pump, a preferred unilateral directionality, and the advent of large scale integration with high yield \cite{pospieszalski2005extremely}. 

At present, state-of-the-art cryogenic HEMT amplifiers are commercially available with $>$ 30~dB gain, a decade of bandwidth, and a noise temperature $T_N \sim$ 1--2~K, in the few 100 MHz \cite{LowNoiseFactory}. These amplifiers typically dissipate $>$~10~mW of power however, which leads to self-heating and a limit on the lowest achievable physical temperature that then also limits $T_N$. Recent work has identified this self-heating as a fundamental challenge since the mW power that is dissipated in the active area of the transistor cannot be removed, even via the use of super-fluid helium \cite{SelfHeating}. Although lower temperatures can be achieved by reducing the transistor bias, this also reduces the transconductance and leads to the amplifier becoming limited by the shot-noise of the drain current \cite{mythesis}. It is interesting that despite the continued improvements in carrier mobility in these heterostructure systems \cite{Manfra}, parametric operation has remained largely unexplored until the present work.


Perhaps one reason high-mobility semiconductor paramps have not been developed relates to the advent of superconducting microwave amplifiers, which have largely addressed the noise limitations imposed by dissipation in hot transistor approaches \cite{arute2019quantum}. Most popular for reading-out the state of qubits are those devices based on variable inductive elements such as Josephson junctions (JJs) or superconductors with large kinetic inductance \cite{aumentado2020superconducting,macklin2015near,mutus2014strong,castellanos2007widely,yamamoto2008flux,white2015traveling,vissers2016low,JoFET}. The power handling capabilities of JJ-based amplifiers  are limited however, by the critical current of the junction, making frequency multiplexing challenging since the total input power is the sum of all frequency channels. Recent efforts may overcome this via the use of kinetic inductance-based travelling wave parametric amplifiers (K-TWPAs) made from 
 NbTiN \cite{eom2012wideband,malnou2020three}, although for semiconductor qubit readout, operation in the few 100 MHz may be more difficult. Significant, of course, is that all superconducting approaches require shielding from magnetic fields and operation below their critical temperature.

Comparing HEMTs and superconducting amplifiers then, we identify a gap in performance between these platforms for applications that require sub-microwatt power dissipation for millikelvin operation, high power handling, and inherent compatibility with magnetic fields [see Table~\ref{tab:AmpComp}]. Indeed, a tightly integrated readout sub-system for semiconductor qubits is such an application, motivating the development of the QCPA which largely fills this gap. Although the noise performance of the QCPA is unlikely to ever rival superconducting approaches, we suggest there is considerable scope to further reduce dissipation leading to improved noise performance.

\section{The QCPA}
\begin{figure}[!h]
\centering{\includegraphics[width=87mm]{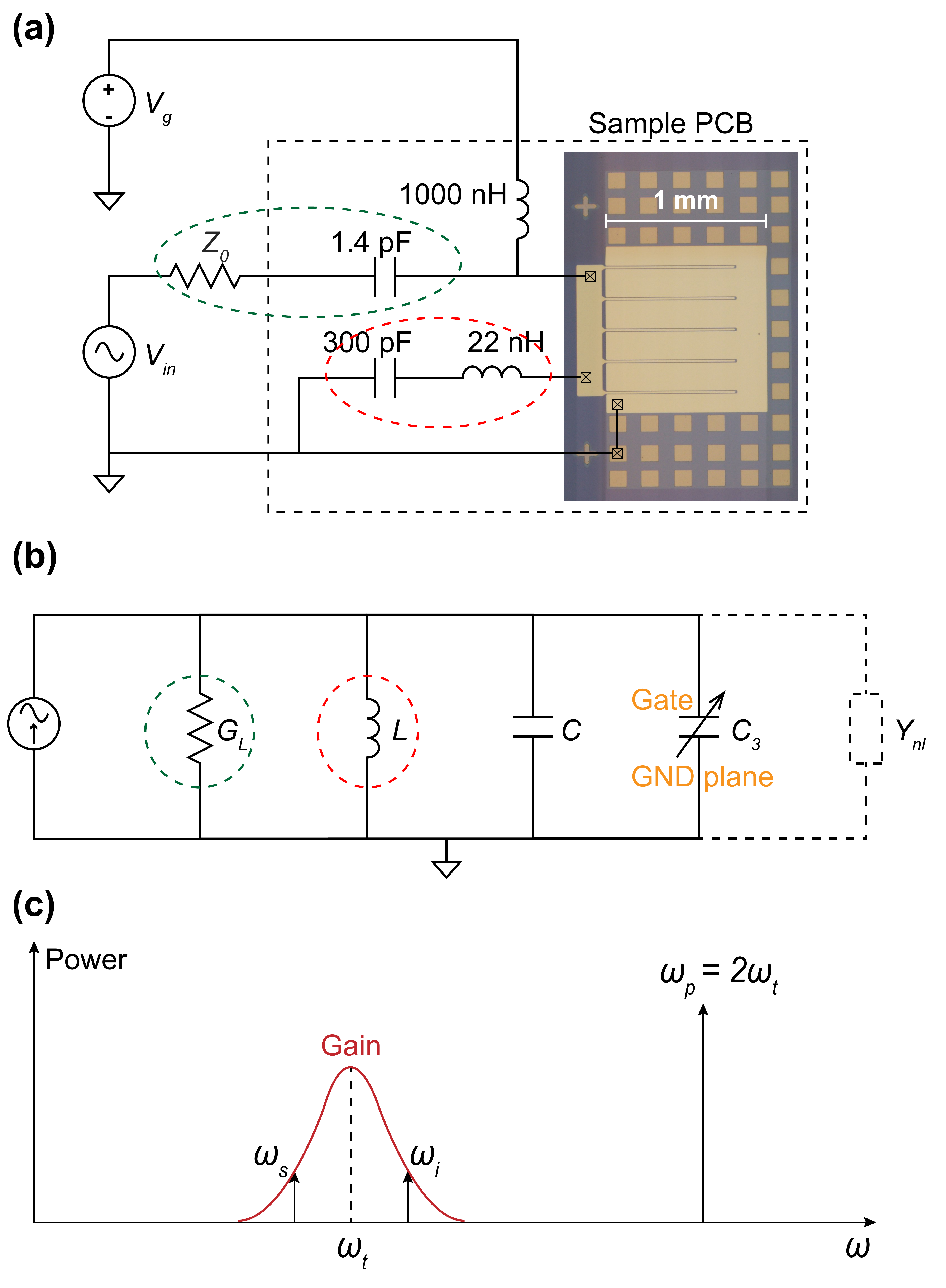}}
\caption{Circuit implementing the QCPA using a gate-configurable 2DEG. (a) Shows a micrograph of the fabricated device with the equivalent parallel tank circuit shown in (b). The equivalence of the various elements is highlighted in (a) and (b) by colored dashed ovals. In (b), the capacitance is divided into two elements, the parasitic capacitances represented by the passive capacitor $C$, and the variable part represented by the variable capacitor $C_3$. The external inductor (22~nH) is in parallel with the active device and forms the second leg of the QCPA tank circuit represented by the inductance $L$ in (b). The coupling capacitance (1.4~pF) transforms the input/load impedance of the circuit to a larger value (series to parallel conversion). The resultant loaded losses of the tank and the source/load are annotated as $G_L$ in (b). The gain and bandwidth are extracted by the analysis of the circuit in (b) by solving for the expression of the equivalent admittance $Y_{nl}$ due to the variable reactance in parallel to the tank circuit at the source frequency. (c) is a spectrum illustration of three-wave mixing. The curve represents the non-degenerate gain with the source and idler signals de-tuned from the center $\omega_t = 1/\sqrt{LC}$ by $\mp\omega_x$ respectively.}
\label{fig:Mix3wave}
\end{figure}

The QCPA is realized via modulating the electron density of a 2DEG formed 91 nm below the surface, at the interface of a GaAs/AlGaAs heterostructure grown using MBE (ungated density = 1.8$\times$ 10$^{11} \text{cm}^2$ and mobility $4.4\times10^6~\text{cm}^2\text{V}^{-1}\text{s}^{-1}$). The variable quantum capacitance is associated with the 2DEG acting as the bottom plate of the capacitor and a 200~nm niobium titanium nitride (NbTiN) metal gate layer acting as the top plate.  A micrograph of the device is shown in Fig.~\ref{fig:Mix3wave}.(a), [see methods section for an illustration of the  device cross-section in Fig.~\ref{fig:ExpSch}]. The top plate of the capacitor is fabricated as a set of five finger gates. Electrical connection to the 2DEG is formed via an array of parallel ohmic contacts comprising an alloy of NiGeAu, with the dc resistance of a single contact being of order 100 $\Omega$. A large NbTiN ground plane surrounds the finger gates [see methods for further details of device fabrication]. As well as being modulated by a pump tone, the quantum capacitance offset is also tuned with a dc voltage applied to the gate via a 1000 nH bias inductor that acts as a block at rf.

To create the resonator tank circuit needed to produce parametric gain, the variable capacitance is bonded to an off-chip, air-core copper spring inductor (22 nH) and then to ground via a 300 pF capacitance, as illustrated in Fig. 1(a). An equivalent parallel circuit implementation is shown in Fig. 1(b) to aid in circuit analysis [see methods for details]. The non-degenerate gain $G_p$ of the amplifier as a function of frequency can be then be written as \cite{mythesis}:
\begin{equation} \label{Eq:Gpmain}
    G_p = \frac{G_{nl}^2+x^2 \left[ G_{nl}+G_L(1+x^2) \right]^2}{\left[ -G_{nl}+G_L(1+x^2) \right]^2+x^2 \left[ G_{nl}+G_L(1+x^2) \right]^2},
\end{equation}
with $G_{nl}$ being the effective negative conductance due to the pumped variable capacitance at the tank resonance frequency $f_t = \omega_t/2\pi$, $G_L$ the total loaded losses, $Q_t$ the loaded quality factor of the amplifier, and $x = 2Q_t\omega_x/\omega_t$ representing the frequency detuning $\omega_x$ of the source signal from the tank resonance $\omega_t$, as illustrated in Fig. 1(c).
Furthermore, at high gain the amplifier has a linear gain-bandwidth relationship:
\begin{equation} \label{Eq:GBWmain}
    \sqrt{G}\times BW = \frac{f_t}{2Q_t}
\end{equation}
with $G$ being the maximum non-degenerate gain of the amplifier at the center frequency and $BW$ is the bandwidth \cite{mythesis}. 

\section{Experimental Results}
Measurements are carried out over a series of cool-downs with the amplifier at the base of a dilution refrigerator. Attenuators on the source tone line are chosen to reduce its noise temperature to a few milli-kelvin. Filtering on the pump and the dc ports removes their respective noise contributions at the source frequency (360--370 MHz) [see methods for full experimental schematic in Fig.~\ref{fig:ExpSch}]. The pump power is not attenuated since amplitudes of order 10s of millivolts are needed at the gate. Note that a modified experimental setup is used for noise characterization, as described below.

\begin{figure}[!h]
\centering{\includegraphics[width=87mm]{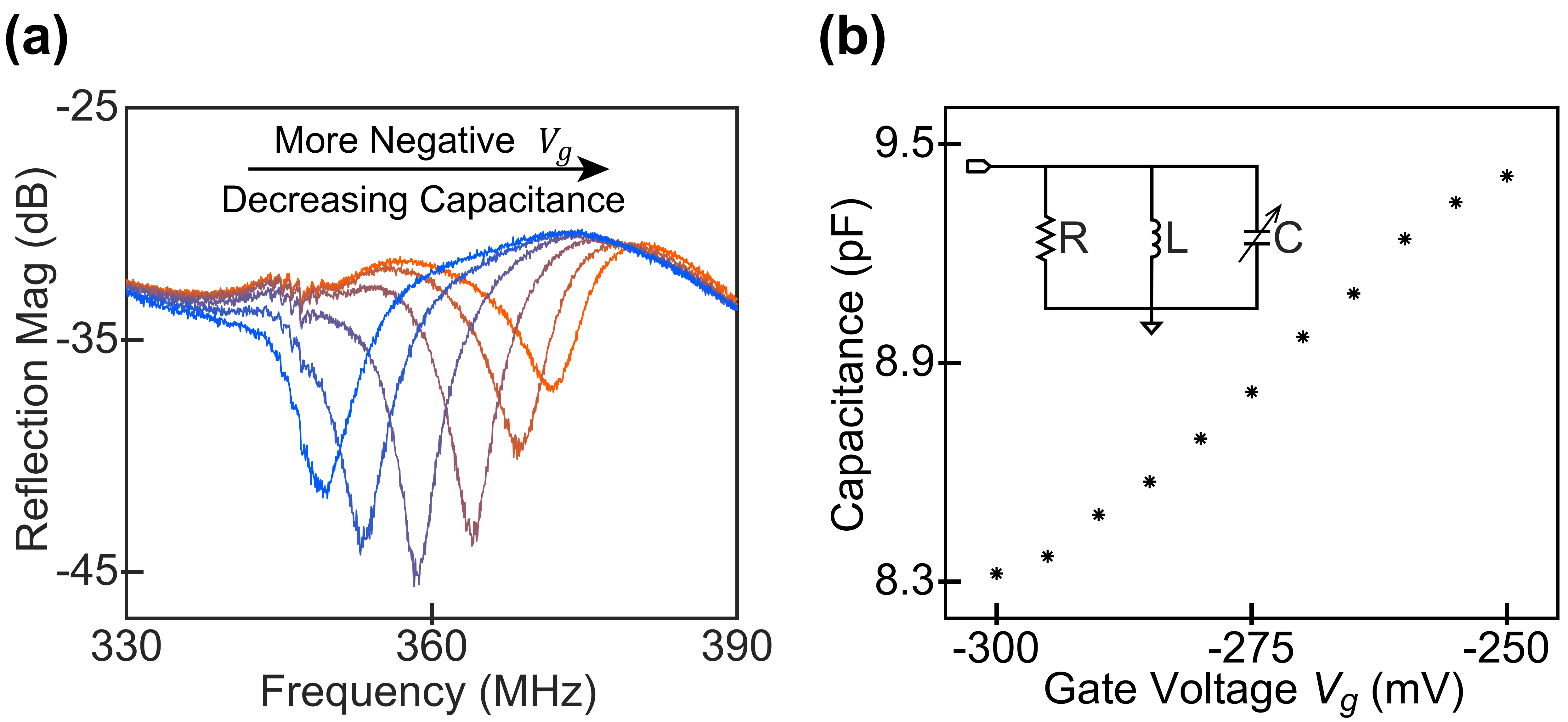}}
\caption{The frequency response of the QCPA is characterized as a function of gate voltage, $V_g$, using a vector network analyzer to apply a low-power input tone. (a) The reflection coefficient magnitude is plotted for $V_g$ ranging between -250~mV and -300~mV. (b) The gate capacitance is extracted from the shift in resonance frequency of the tank circuit (shown in the inset).}
\label{fig:CV}
\end{figure}

The reflection coefficient of the QCPA is measured as a function of frequency at low source power ($-100~\text{dBm}$) as the dc voltage applied to the gate is made more negative, as shown in Fig.~\ref{fig:CV}.(a). The shift in resonance frequency with gate voltage indicates a tunable capacitance, which can be determined from the resonance frequency using $f_t = 1/2\pi\sqrt{LC}$ as shown in the simplified tank circuit in the inset of Fig.~\ref{fig:CV}.(b). Here the capacitance is plotted as a function of gate voltage, where the slope of the C--V characteristic gives a lever arm of $26.7~\text{pF/V}$.

\begin{figure}[!t]
\centering{\includegraphics[width=87mm]{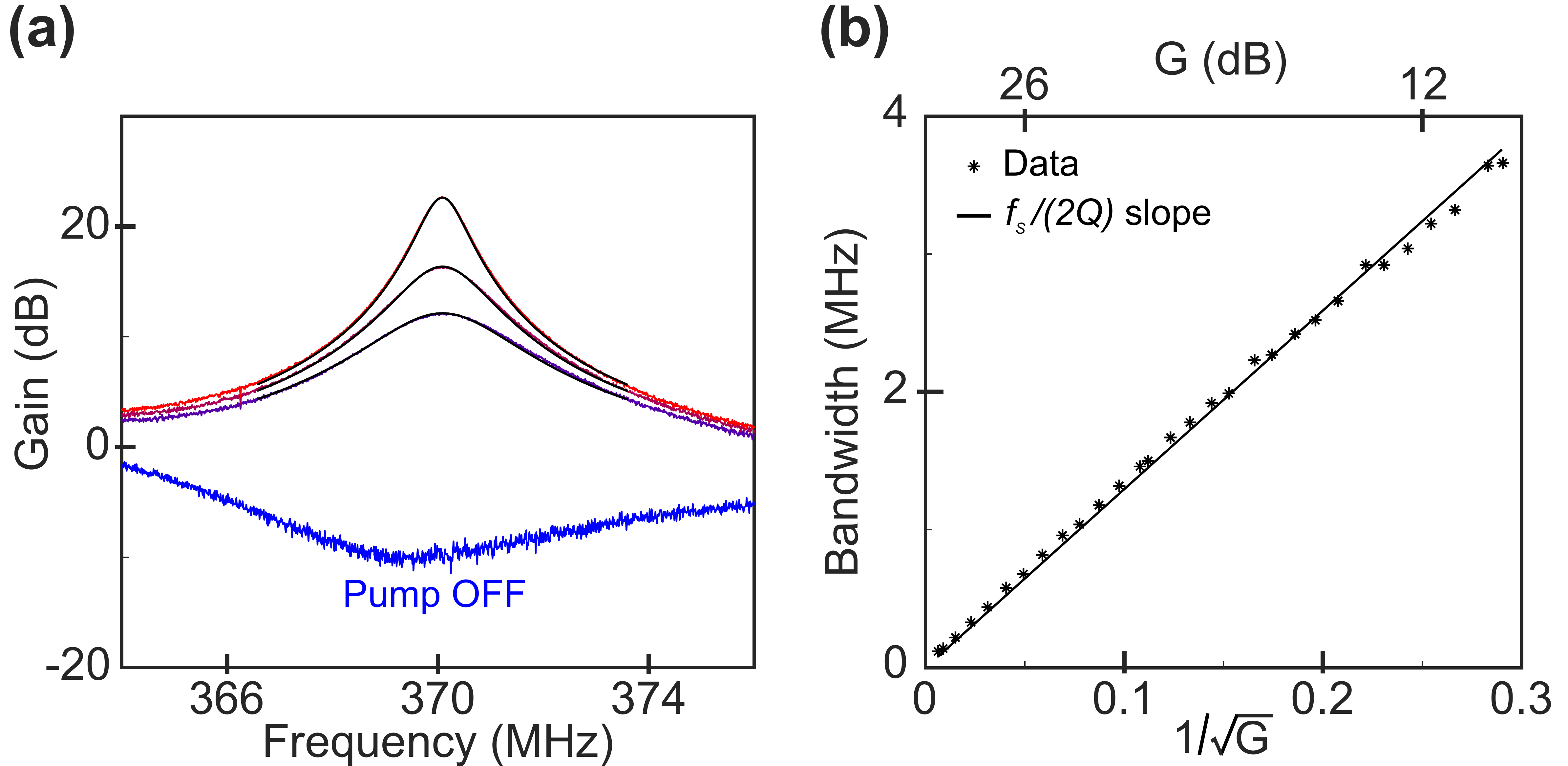}}
\caption{The gain of the QCPA is measured using a network analyzer while the device is pumped at $\omega_p = 740.17~\text{MHz}$. (a) The gain is plotted as function of frequency and pump power, 
$P_p=-26.4~\text{dBm}, -25.6~\text{dBm}, -24.8~\text{dBm}$ respectively. 
The overlaid black curves show the gain predicted by the theoretical model of Eq.~\ref{Eq:Gpmain}. (b) The bandwidth of the amplifier is extracted and plotted as a function of the inverse square root of the gain, verifying the relationship in Eq.~\ref{Eq:GBWmain}.}
\label{fig:GBW}
\end{figure}

\begin{figure}[!b]
\centering{\includegraphics[width=87mm]{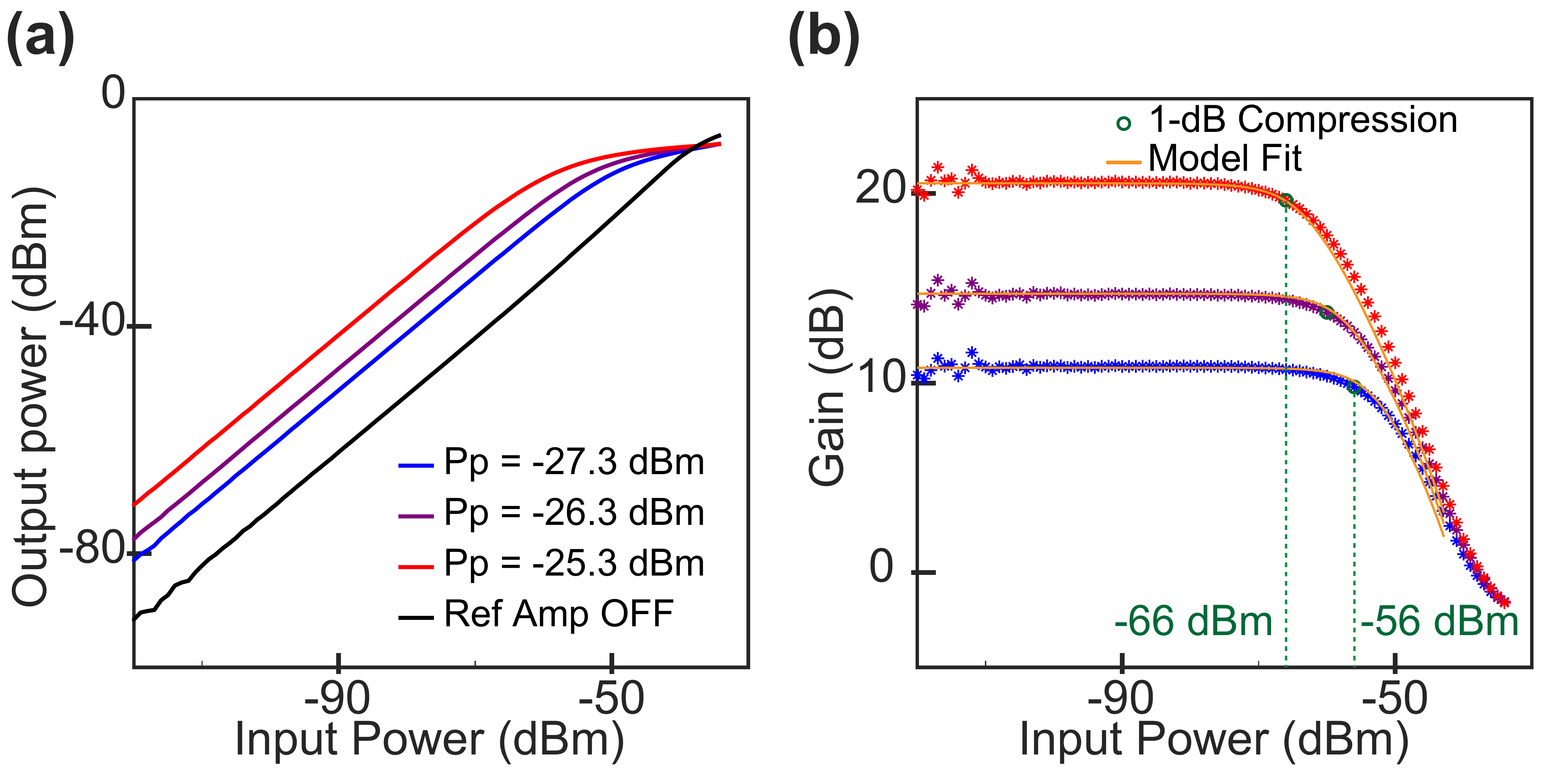}}
\caption{The QCPA non-degenerate gain is measured as a function of source signal power. (a) shows the output power measured at the spectrum analyzer as a function of the input power referred to the input of the QCPA. A reference curve is plotted in black for the case when the amplifier is OFF. (b) The gain of the QCPA is plotted as a function of increasing input power with the theoretical saturation behaviour overlaid (gold coloured lines). }
\label{fig:DR}
\end{figure}

We next look for amplification by varying the pump power and source frequency. The gain of the amplifier manifests as an increase in reflected signal power when  pumping, relative to the idle state (when the pump is OFF and dc gate bias set to $V_g $= 0 V). Measured gain as a function of frequency is plotted in Fig.~\ref{fig:GBW}(a), for different pump powers. Here the tank circuit center frequency is $370.085~\text{MHz}$ and the pump is at $2\times370.085~\text{MHz}=740.170~\text{MHz}$. We fit the gain model using Eq.~\ref{Eq:Gpmain}, with $Q_t = 14.3$ and a pumped capacitance of $C_3 = 0.51~-~0.57~\text{pF}$. Plotting the amplifier bandwidth with $1/\sqrt{G}$ leads to the linear dependence shown in Fig.~\ref{fig:GBW}(b), typical of parametric operation \cite{mythesis,schackert2013practical} as described by Eq.~\ref{Eq:GBWmain}. Measuring the input power at which the amplifier saturates, Fig.~\ref{fig:DR}(a) and (b) indicates 1-dB gain compression is reached at input powers of -66~dBm and -56~dBm at gains of 20.6~dB and 10.8~dB respectively.

\begin{figure}[!h]
\centering{\includegraphics[width=87mm]{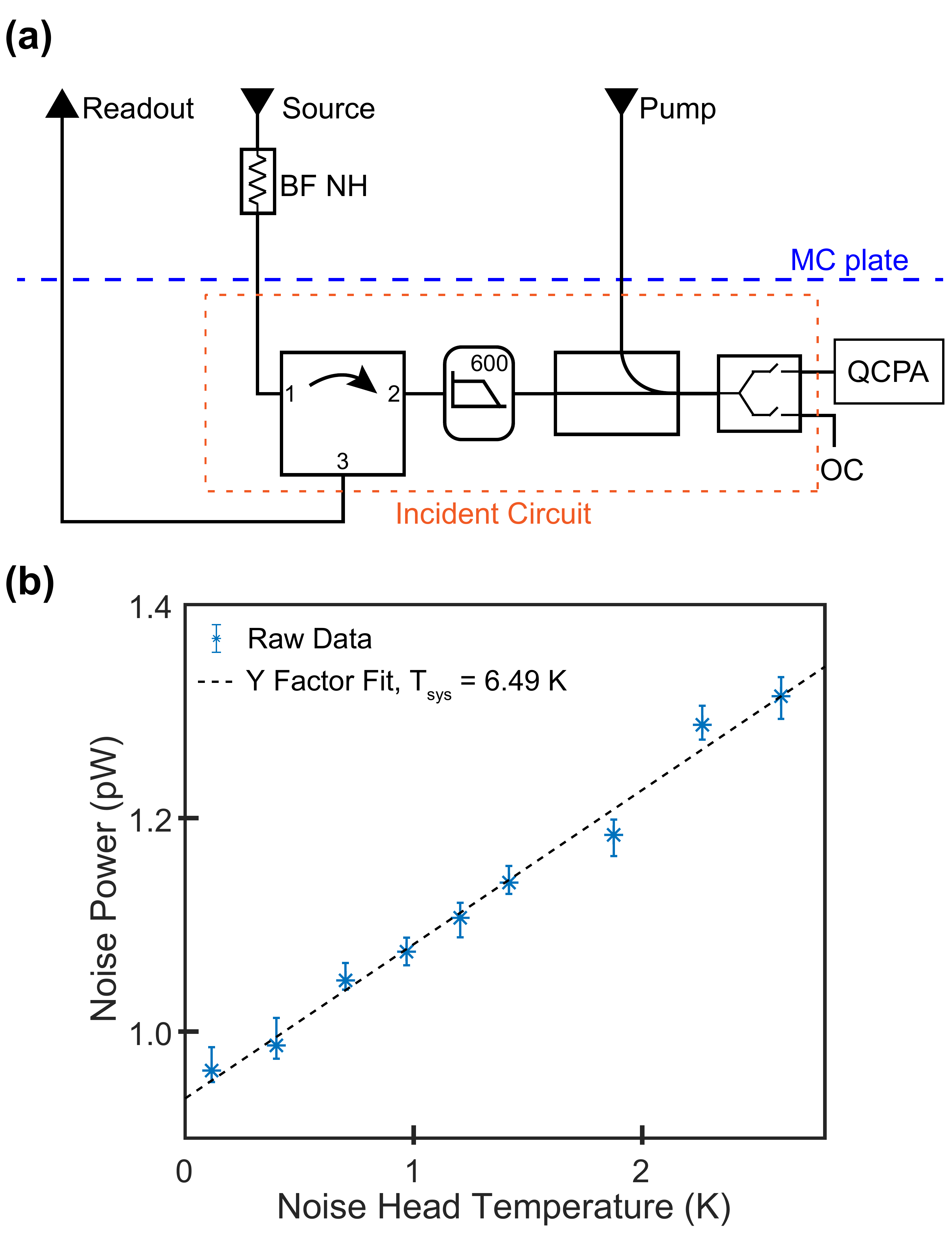}}
\caption{The setup used for noise characterization and measurement. (a) shows the circuit of the QCPA below the mixing chamber plate of the dilution refrigerator. (b) The measured system noise power as a function of noise source temperature along with the corresponding linear fit.}
\label{fig:NoiseHead}
\end{figure}

Turning now to the noise performance of the QCPA we modify the setup to incorporate a commercial cryogenic noise source \cite{simbierowicz2021characterizing}, as shown in Fig.~\ref{fig:NoiseHead}(a). The pump signal is coupled to the input of the amplifier and a circulator routes the reflected signal to the readout line. A cryogenic switch is also added to the setup in order to compare and normalize the amplifier noise response relative to the response of an open circuit [labeled ``OC" in Fig. 5(a)]. Measuring the integrated noise power at the readout port as a function of noise input power then yields the offset noise in the limit where the input is zero. Using this (Y-factor) method, the noise temperature of the system is determined to be 
$T_{sys} = 6.49^{+1.06}_{-0.62}$~K with QCPA gain  $G_{QCPA} = 33.86$~dB.
The bounds on $T_{sys}$ are given by the extrema of the Y-factor fitting within the measured error bars of the noise floor.

In comparison, with the switch [Fig.~\ref{fig:NoiseHead}(a)] positioned to the open circuit port, the system noise temperature is measured to be 48.73~K/$G_{QCPA}$ $\sim$ 20 mK, thus the additional noise beyond the QCPA is negligible and $T_{sys} \sim T_{QCPA}$. This value however, comprises both source tone and folding of the idler noise. In order to subtract the idler noise \cite{simbierowicz2021characterizing} we measure its noise temperature and gain to be $6.40$~K and $33.90$~dB respectively, almost identical to the source signal. Therefore, the noise contribution at the source and idler can be considered equal and the noise temperature of the QCPA at the source only is found as $T_{sys,source} = 3.25^{+0.53}_{-0.32}$~K. 

Finally, it is also necessary to account for transmission losses between the noise source and the input of the amplifier, which serve to increase the effective measured system noise [see methods section for details]. We estimate these losses by independently measuring the calibrated transmission path in a separate cooldown. Owing to the poor performance of the ferrite circulator at these temperatures we find the transmission losses to be $\sim$ 4.01~dB. Correcting for these losses we estimate the intrinsic noise temperature of the QCPA to be $T_{QCPA} = 1.29^{+0.21}_{-0.13}$~K.

It is interesting to compare this measured noise temperature to the estimated physical temperature of the amplifier during pumping. To do this we raise the base temperature of the refrigerator, finding that the amplifier noise temperature only increases when the physical temperature exceeds $\sim$ 1 K. This suggests that the present noise performance is limited largely by the high physical temperature of the device arising from self-heating (with opportunities for improvement are discussed below).

\begin{figure}[!h]
\centering{\includegraphics[width=87mm]{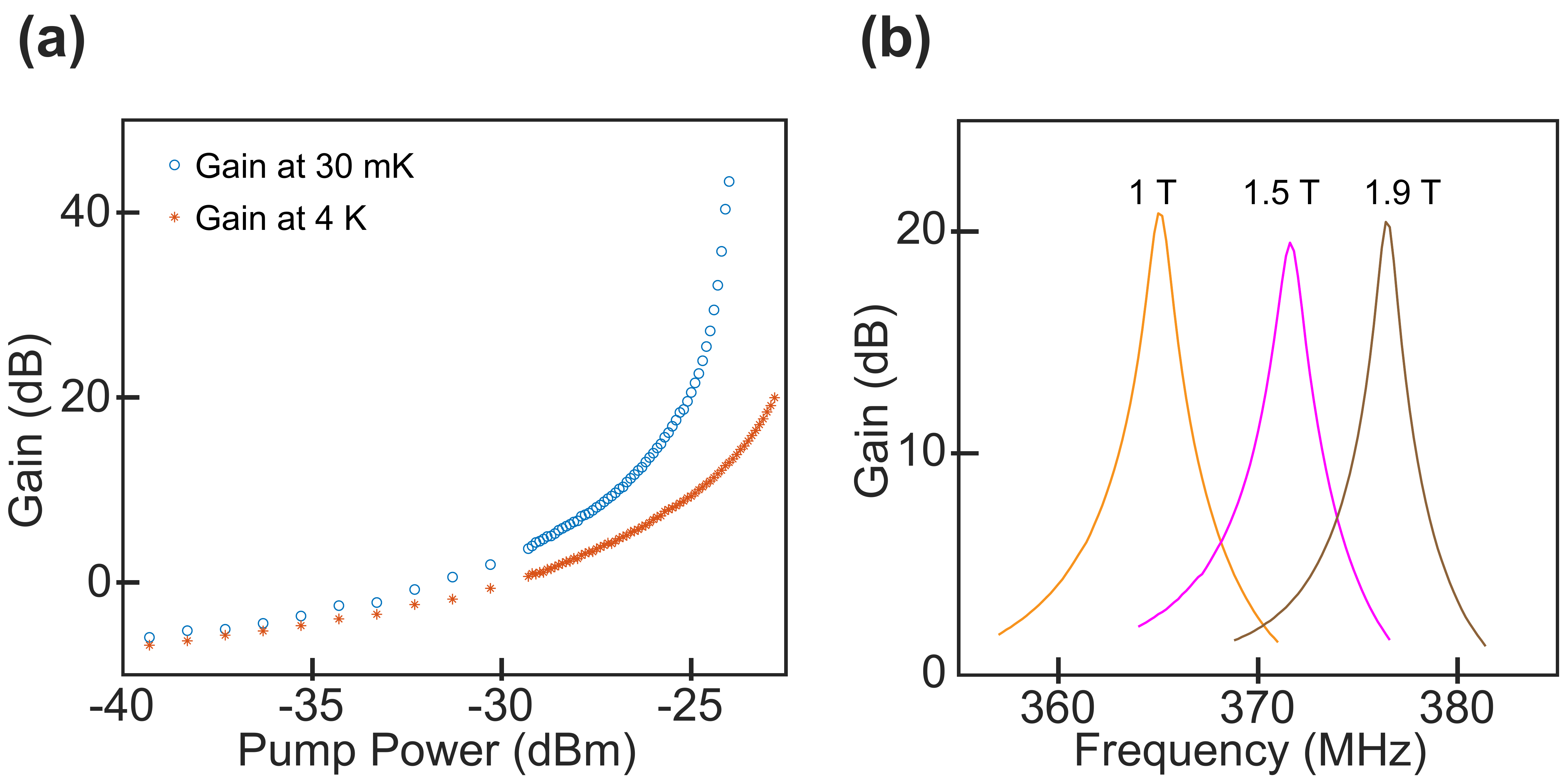}}
\caption{(a) The QCPA is tested at $T\sim$ 4~K with gain  shown as a function of pump power. The $T\sim$ 0.03~K gain profile of the amplifier is shown for comparison. (b) The QCPA gain versus frequency, plotted at 3  magnetic field values (1.0, 1.5, and 1.9~T). The amplifier achieves 20~dB gain with the operation point optimized accordingly, the pump power is set to -23.6~dBm, -17.4~dBm, and -12.3~dBm respectively.}
\label{fig:MF4K}
\end{figure}

Lastly we explore operation of the QCPA in regimes of high temperature and high magnetic field. Raising the cryostat temperature from 30 mK to 4 K we observe a reduction in the maximum gain, particularly evident at high pump power, as shown in Fig. 6(a). Given that resistive losses are largely temperature independent below 4 K the reduction in gain is likely due to a decrease in the transconductance with thermal broadening (as well as thermal population of higher sub-bands in the 2DEG) and partly caused by the loss of superconductivity of the Al bond-wires within the amplifier's tank circuit. Nevertheless, we note that the QCPA yields a gain in excess of 20~dB at elevated temperature, enabling troubleshooting and deployment of a readout sub-system at 4 kelvin (or even higher temperatures). 

Turning to the behaviour of the amplifier in a magnetic field, we note that various qubit platforms require tesla-scale fields for operation, usually oriented in-plane with the 2DEG (as we have here). With the amplifier at 30 mK, we continue to observe moderate gain ($>$20 dB) up to fields of several tesla, as shown in Fig. 6(b). Interestingly, the data also reveal a small reduction in the quantum capacitance of the 2DEG with parallel field, which leads to an increase in resonance frequency. This change in capacitance can be related to the well-known dependence of the 2D density-of-states with field \cite{BEENAKKER19911}.

\section{Discussion and Conclusion}
Having presented experimental results benchmarking our prototype device, we now discuss means of improving its performance as well as touching on additional parameters of the amplifier that are critical for its use in a scaled-up quantum computing or receiver system. 

Starting with the opportunities to lower the noise temperature of the amplifier, we note that this is currently limited by the large (mV) pump amplitude and the presence of lossy circuit components, which together lead to self-heating and a correspondingly high physical temperature (near 1 kelvin). Reducing the pump power by more than 20 dB appears relatively straightforward via the use of a tank circuit to provide impedance matching on the pump port. Further reduction in pump power can be achieved by increasing the C-V lever-arm by moving the 2DEG closer to the surface gate during the growth of the quantum well. 

There are also several pathways to lowering dissipation. For the present device, losses in the normal metal (off-chip) inductor as well as losses in the ohmic contacts likely dominate dissipation. Both of these contributions can be improved by increasing the thickness and width of metal and optimization of alloying and annealling conditions. Alternatively, as noted earlier, moving from GaAs to  InAs (or similar III-V compounds) would lift the challenge of making low resistance ohmic contacts since Fermi-level pinning in such materials leads to the absence of a Schottky barrier at the metal-semiconductor interface. On-the-other-hand, GaAs will likely remain superior in terms of its accessible mobility and ease of fabrication. A further step in reducing dissipation could also come from incorporating superconductivity in the contacts \cite{supersemi}, or in the 2DEG itself via the proximity effect \cite{InAs_SC}. Ultimately the noise temperature will be limited by the electron temperature of the 2DEG in the limit of high pump power (likely a few 100 mK). 

Although these improvements can reduce the noise temperature by lowering the physical temperature, we note that power dissipation in the present device is already sufficiently small to enable operation at milli-kelvin temperatures. We estimate that this dissipation is of order a few 100 nW, commensurate with 100s of readout sub-systems operating within the cooling power of a commercial dilution refrigerator at 100 mK.


It is also worth considering the current size of our prototype device, which is of order a few $\text{mm}^2$. This footprint can likely be reduced significantly via the use of shallower 2DEG to maintain a sizeable gate capacitance. In such a configuration 100s of amplifiers can be fabricated on a single centimeter-scale chip, that is either monolithically integrated with 2DEG-based qubits, or positioned proximal to the qubit plane using heterogeneous chip-to-chip interconnect approaches. 

We also draw attention to the likelihood that the QCPA can be configured as a travelling-wave device, where  the quantum capacitance element is integrated in a matched coplanar waveguide to achieve wideband operation, similar to what has been achieved in superconducting implementations \cite{eom2012wideband,malnou2020three,vissers2016low,macklin2015near}. The use of multiple pump tones with fixed phase offsets would also see the QCPA exhibit parametric non-reciprocity, enabling rf circulators and isolators to also be integrated on-chip in this 2DEG platform.

Stepping back from the specific details of our implementation, the use of a voltage controllable, low-loss, capacitance as a parametric element in which to realize amplification is a general result that applies broadly to a spectrum of nanoscale devices exhibiting a sizable quantum capacitance. Examples include other 2D systems based on silicon MOS structures \cite{West}, silicon germanium heterostructures \cite{HRL,meno}, van der Waals (vdW) heterostructures and graphene\cite{grapheneothers}, or even 1D devices \cite{Hu,MCJ}. Considering that many qubit platforms already leverage the dispersive response of an rf resonator for readout\cite{colless2013dispersive,West}, there is also the potential to combine these readout resonators \cite{hornibrook2014frequency} with parametric capacitive elements that provide $in-situ$ amplification directly. 

In conclusion, we have realized a prototype parametric amplifier based on gate-modulation of a high mobility 2DEG in a semiconductor heterostructure. With considerable opportunity for further refinement, the specifications of our amplifier likely fill a gap between the lowest noise superconducting amplifiers and high-power transistor amplifiers. We anticipate the QCPA may find utility in realising low power, integrated readout sub-systems for semiconductor qubits, deep-space receivers, and radio astronomy applications. 

\section{Methods}
\small{
\subsection{Sample Fabrication} 
NiGeAu contacts are deposited first on the GaAs surface and annealed at 450 C. The active area of the device is defined by a mesa etch with a subsequent 16 nm-thick layer of $\text{Al}_2\text{O}_3$ grown by atomic layer deposition (ALD). This layer enables operation beyond the gate bias defined by the Schottky barrier, passivating the semiconductor surface and suppressing leakage. In the last step of the fabrication, a metallic gate and ground plane (NbTiN) structure is patterned with optical lithography. The actual device photographed in Fig.~\ref{fig:Mix3wave} comprises a dummy TiAu gate and ground plane (instead of NbTiN), but was not electrically characterized. With this exception it is otherwise identical to the device characterized. 


\subsection{Dissipation and Characterization} 
Losses are mitigated via two approaches. Firstly, a large number of ohmic contacts are used in parallel (with resistance measured to be of the order of $100~\Omega$ per contact at dc). The relatively large contacts ($100\times100~\mu\text{m}^2$) will also couple capacitively to the 2DEG reducing the impedance significantly at radio frequencies. Secondly, the ground plane is deposited relatively close to the gate ($7.5~\mu\text{m}$ away) and serves to provide a low-impedance path to ground at high frequencies by forming a large capacitance (order of nF) with the active layer of the 2DEG. 

The variable capacitance of the active device is extracted as a function of gate voltage in the course of characterizing the QCPA in Fig.~\ref{fig:CV}.(b). The overall capacitance change is $\approx 280~\text{fF}$ over a voltage range of $10~\text{mV}$. For the purpose of the QCPA acting as a 3-wave mixing circuit, the desirable C-V characteristic is one with a linear component in voltage. A linear dependence of the active capacitance on voltage ($C_v = C_c v$ where $C_c$ is the constant slope in [F/V]) introduces the mixing of the pump with the idler and source signals resulting in their amplification. To enhance the linearity of the C-V curve, we chose a gate layout with an extensive perimeter. 

\begin{figure}
\centering{\includegraphics[width=70mm]{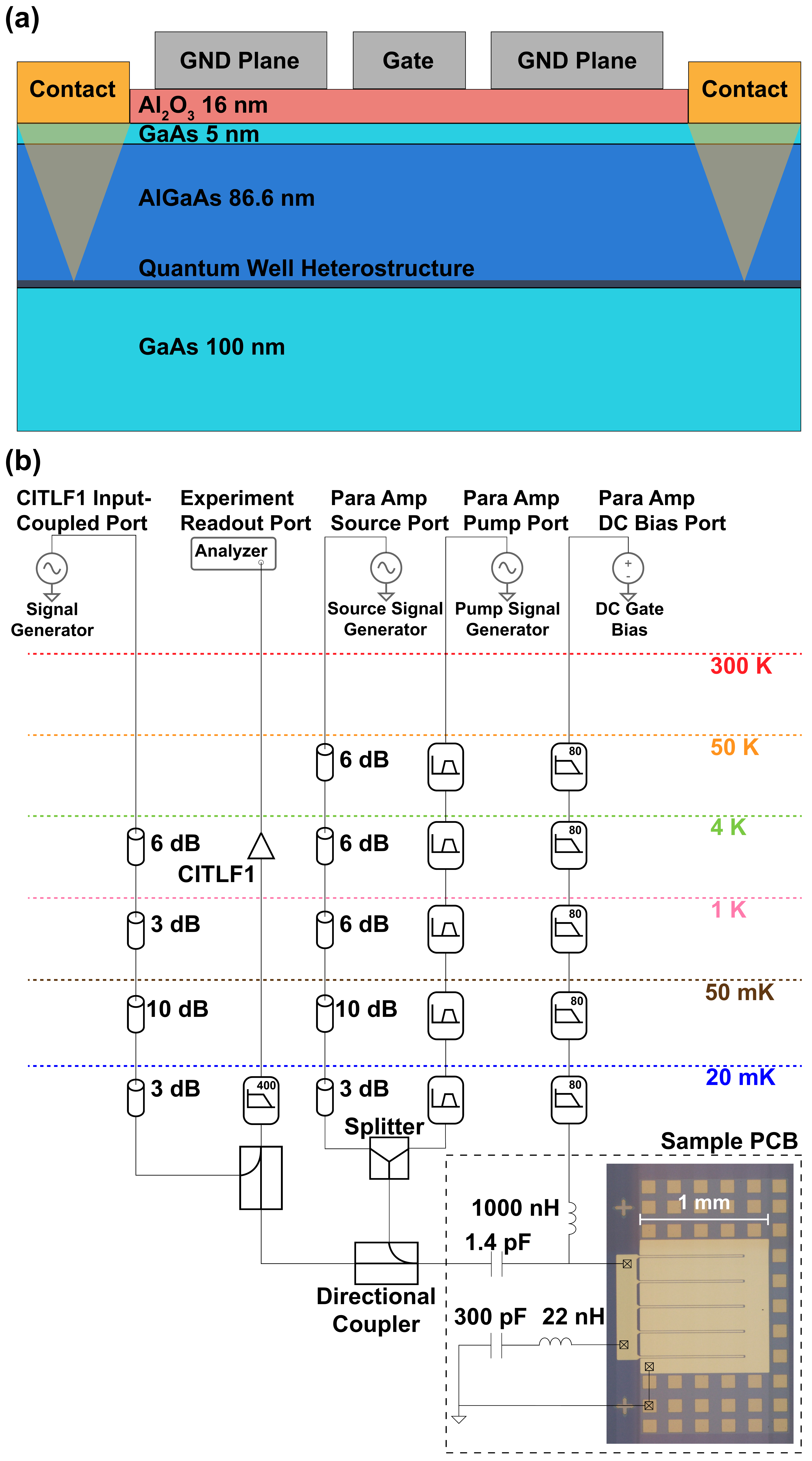}}
\caption{(a) An illustration of the device cross section (not to scale).  (b) The experimental setup used to characterize the QCPA. The source signal port is heavily attenuated to reduce the noise temperature in the band of the amplifier to the mK range. The pump port is filtered to pass only the pump signal and sharply reject signals in the input/idler band. This is to prevent noise in the pump signal at the source frequency from corrupting the system. The dc bias port is low-pass filtered at 80 MHz. The readout signal is amplified at 4~K plate of the refrigerator with a commercial Silicon-Germanium low-noise amplifier CITLF1. A 400 MHz low-pass filter prevents the high-powered pump signal from saturating the 4~K low noise amplifier (LNA). An additional port is configured to couple signals into the 4~K LNA independent to the QCPA. It is intended to debug/characterize the 4~K LNA if needed. The PCB is highlighted by a dashed block below the coldest stage.}
\label{fig:ExpSch}
\end{figure}
\subsection{The QCPA circuit} \label{app:QCPAcircuit}
The device is fixed to a printed circuit board using silver paste and bonded to the circuit shown in Fig.~\ref{fig:Mix3wave}.(a) forming the QCPA tank. The Ohmic contacts and the ground plane of the active device are bonded to the ground plane of the PCB. The tank inductance is an air-core spring inductor of 22~nH connected in series with a 300 pF capacitance to ground. This capacitance acts as a dc block for the gate bias, but is negligible at high frequency. The gate bias is applied through an rf choke inductance of $1~\mu\text{H}$. The coupling capacitance of 1.4~pF serves to transform the impedance seen by the amplifier at the load from $50~\Omega$ to $1.9~\text{k}\Omega$ at the frequency of interest (370~MHz). This effectively raises the loaded Q-factor of the QCPA roughly by a factor of 19 times. It also forms a reasonably good match with the amplifier tank so its resonance can be easily seen as in Fig.~\ref{fig:CV}.(a) allowing characterization of the device further. The coupling capacitor along with the choke inductor form a bias tee on the PCB. The experimental circuit implemented on the PCB is then equivalent to a parallel tank parametric amplifier circuit which can be analytically solved for the gain as shown in Fig.~\ref{fig:Mix3wave}.(b).

\subsection{Degenerate Gain} \label{app:DegGain}
\begin{figure}
\centering{\includegraphics[width=70mm]{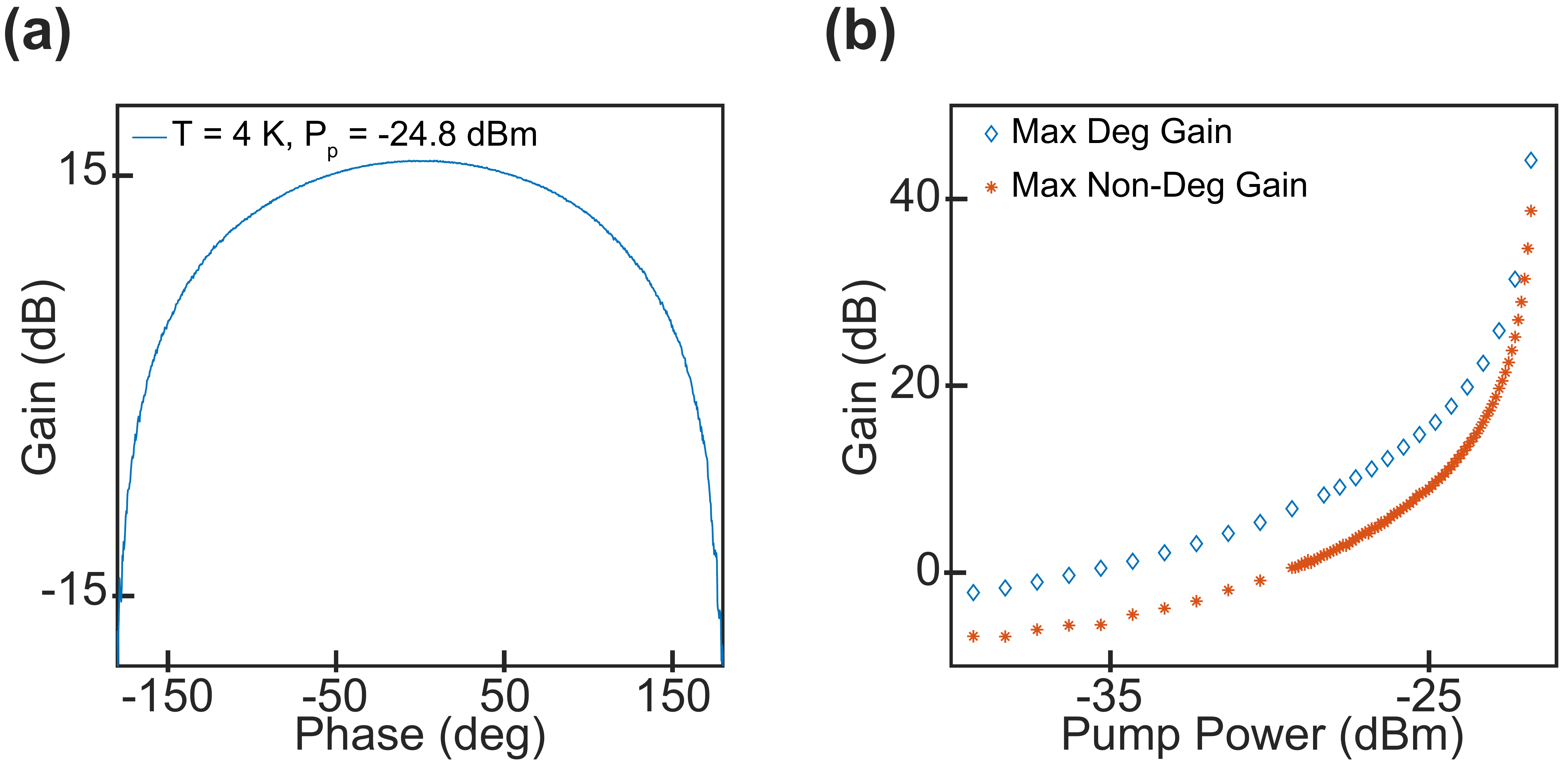}}
\caption{The degenerate gain of the parametric amplifier was characterized at 4~K, plotted in (a) as a function of the pump phase. (b) shows a comparison between the maximum attainable degenerate gain and non-degenerate gain.}
\label{fig:DegG}
\end{figure}
The maximum gain attained by the QCPA is the degenerate gain where $f_s = f_i = f_p/2 = f_t$ with a certain phase arrangement between the pump and the source tones. On the other hand, with a different phase arrangement, $Y_{nl}$ could take a real positive value which then acts as an attenuator of the source signal. Thus, the operation of the amplifier in the degenerate case is phase-dependent and can swing from high amplification to deep attenuation. Fig.~\ref{fig:DegG} shows the characterization of the degenerate gain at a temperature of 4~K.

\subsection{Noise Temperature Correction and Calibration} \label{app:NoiseTemp}
In order to extract the noise temperature of the QCPA individually, the experiment consists of measuring the spectra of the QCPA and its reference open-circuit termination with a sweep of the cryogenic noise source physical temperature. The full schematic of the experiment is illustrated in Fig.~\ref{fig:NoiseSchem}. To ensure accurate reading of the noise floor, we set the spectrum analyzer (Keysight N9030B) to capture with a resolution bandwidth of 200~Hz, a video bandwidth of 1~Hz, and video averaging of 500 curves. This ensures that the error bars in the noise measurement are kept within reasonable margins. The frequency span is 10~kHz centered around the degeneracy center frequency of the amplifier. An example of the raw spectrum data at base temperature $T_{amb}=12$~mK is shown in Fig.~\ref{fig:NoiseElevTemp}.(a). The noise power is extracted as the median value of the noise floor in the source and idler side-bands of the amplifier and the Y-factor method is used to extract the noise temperature of the corresponding system. 

\begin{figure}
\centering{\includegraphics[width=87mm]{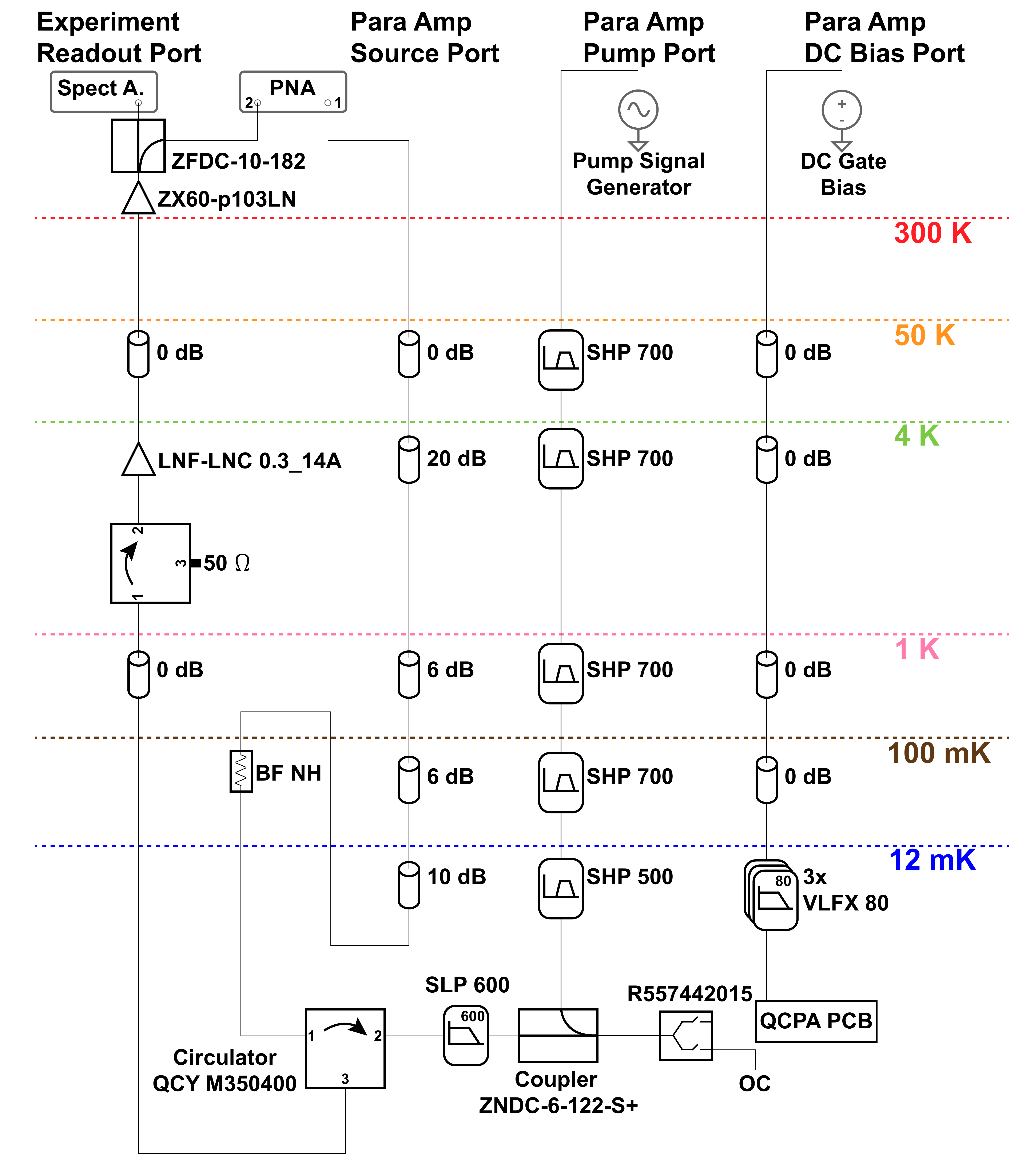}}
\caption{Schematic of the noise measurement experiment which is largely similar to the original setup in Fig.~\ref{fig:ExpSch}.(b) with the addition of the cryogenic noise source to the setup and the use of circulators in place of the couplers.}
\label{fig:NoiseSchem}
\end{figure}

\begin{figure}
\centering{\includegraphics[width=87mm]{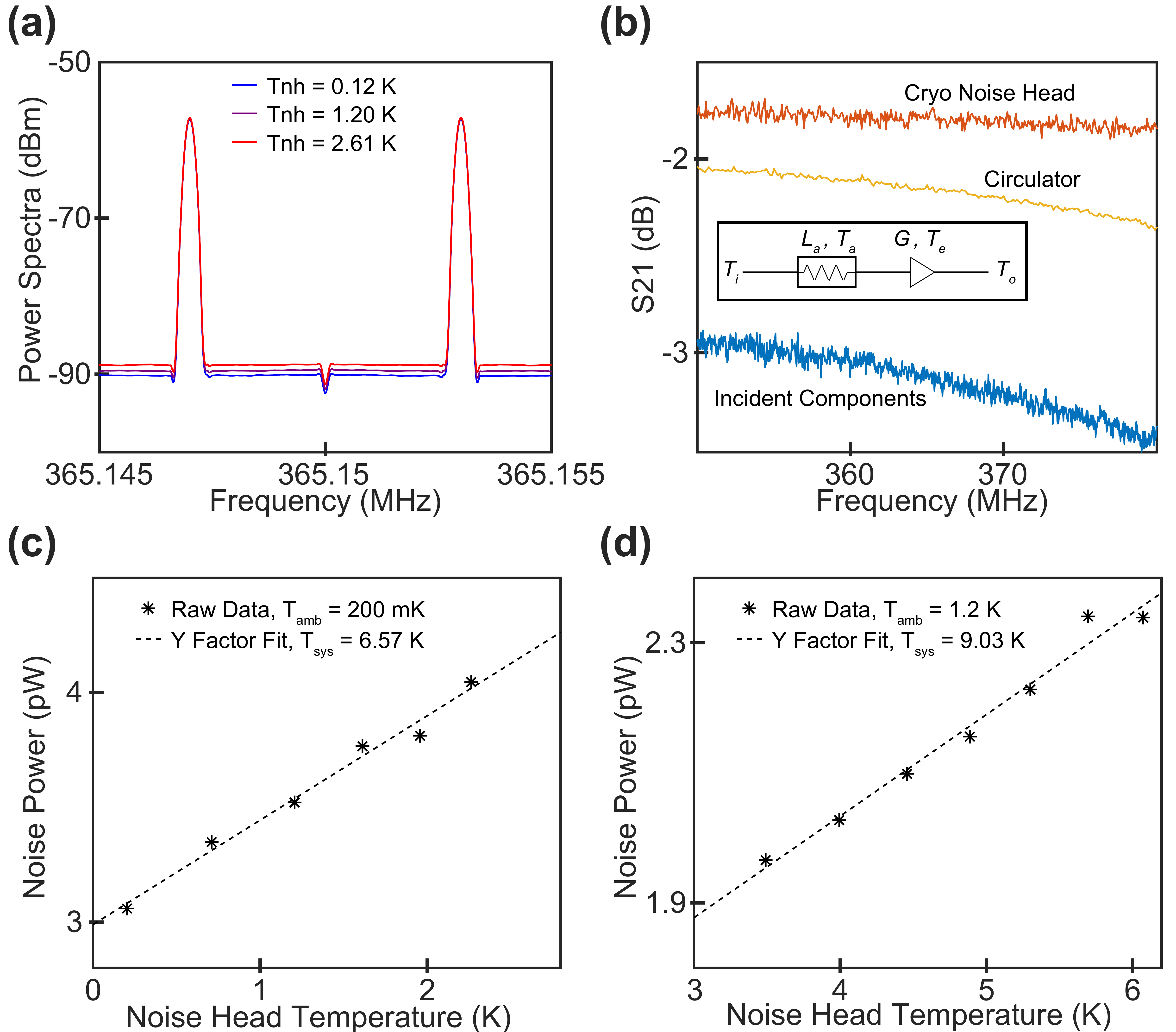}}
\caption{(a) A sample of the raw spectra used for the computation of the noise temperature at $T_{amb} = 12$~mK. The legend shows the physical temperature of the noise head for each curve. (b) The calibrated $S_{21}$ measurements of the incident components (circulator, the low-pass filter, the directional coupler, and the switch) all kept in the same assembly as used in the main noise experiment of Fig~.\ref{fig:NoiseHead}, the Cryo noise head excess (above 20~dB) insertion loss, and the individual circulator QCY-M350400. (c-d) The QCPA system noise temperature is characterized at higher ambient temperatures by applying heat to the mixing chamber stage: (c) $T_{amb} = 200$~mK. (d) $T_{amb} = 1.2$~K.}
\label{fig:NoiseElevTemp}
\end{figure}

Furthermore, the incident circuit between the noise head and the QCPA must be characterized. We define the incident circuit as the collection of  components that constitute the signal path for the wave travelling from the output of the noise head (a controllable heated attenuator) down to the input port of the QCPA sample PCB. We place this circuit in a separate dedicated SOLT (short-open-load-thru) calibration setup in a 5~K environment. The SOLT calibration setup consists of a Fairview kit connected in between 2 rf Radiall SP6T (single-pole 6-throw) R594F43617 relay switches where the DUT is connected with same-electrical-length cables as the calibration standards. The entire setup is mounted on the 5~K stage of a cryogenic system. 

The calibration setup is used to measure the loss in the lengthy cables of the noise head assembly individually. In the main noise experiment, the noise head is used as a well-calibrated 20~dB attenuator whose temperature can be controlled via a heater and a temperature sensor. It allows `thru' measurements where the signal enters the noise head, gets attenuated and also acquires a noise temperature equivalent to the physical temperature of the noise head. Due to its potentially elevated body temperature, the noise head is mounted at the 100~mK cold plate (just above the mixing chamber in a BlueFors LD fridge) and is connected to the setup via lengthy superconducting cables to limit thermal conduction. The calibration setup allows measurement of the losses in said superconducting cables. The extra insertion loss above 20~dB is plotted in Fig.~\ref{fig:NoiseElevTemp}.(b). We attribute this loss to the cables where it is divided equally between the input and output sections. The incident circuit includes the output cable of the noise head whose loss is then found to be 0.893~dB at the frequency of interest. 

Next, we measure the rest of the incident circuit (highlighted in Fig.~\ref{fig:NoiseHead}.(a)), i.e the circulator, the low-pass filter, the directional coupler, and the switch all kept in the same assembly as used in the main noise experiment. The $S_{21}$ transmission curve is shown in Fig.~\ref{fig:NoiseElevTemp}.(b). At the frequency of interest, we find that the insertion loss of the above components to be 3.113~dB. The total insertion loss of the incident circuit is then found to be 4.006~dB.  

The correction of the system noise temperature due to the incident circuit losses is detailed in the following. Consider the schematic in the inset of Fig.~\ref{fig:NoiseElevTemp}.(b) where the input noise temperature is $T_i$, the amplifier is characterized by a gain $G$ and noise temperature $T_e$, and the attenuation is characterized by a loss $L_a$ and physical temperature $T_a$. Through this circuit, we can investigate the effect of the existence of attenuation $L_a$ on the noise temperature of the system. The characteristic noise temperature of the attenuation is related to its physical temperature and defined as $T_a(L_a-1)$. 
 The noise temperature at the output of the attenuator is then written as:
 \begin{equation}
     T_{o,a} = \frac{T_i+T_a(L_a-1)}{L_a}
 \end{equation}
The output noise temperature $T_o$ of the whole system can be expressed as:
\begin{equation}
    T_{o} = \frac{G}{L_a}T_i + \frac{G}{L_a}T_a(L_a-1) +GT_e.
\end{equation}
Defining the total gain of the system as $G_t = G/L_a$, the system noise temperature is $T_{sys} = T_o/G_t - T_i$:
\begin{equation}
    T_{sys} = T_a(L_a-1) +L_aT_e.
\end{equation}
In the system of the QCPA noise measurement experiment where $L_a = 2.52$ (4~dB) and assuming the temperature $T_a = 100$~mK, the term $T_a(L_a-1) = 152$~mK is a very small correction compared to the measured $T_{sys} = 6.49$~K. However, the more pronounced effect of $L_a$, which we aim to correct for, is the term $L_aT_e$ where it multiplies the noise temperature of the QCPA by the losses of preceding incident circuit causing the system noise temperature to increase by $2.52$ times.


}
\section{Acknowledgements}
We thank S. Pauka, J. Colless, J. Hornibrook, T. Ohki, M-C. Jarratt, D. Govender, S. Waddy, J. Witt, A. Jouan, and K. Zuo for stimulating conversations and technical contributions. This research was supported by the Microsoft Corporation and the Australian Research Council Centre of Excellence for Engineered Quantum Systems (EQUS, CE170100009). The authors acknowledge the facilities as well as the scientific and technical assistance of the Research \& Prototype Foundry Core Research Facility at the University of Sydney, part of the Australian National Fabrication Facility.

$\dagger$ Corresponding author: David.Reilly@sydney.edu.au 
\bibliographystyle{apsrev4-2.bst}
\bibliography{apssamp}

\end{document}